\documentclass[a4paper]{jpconf}
\usepackage{graphicx}
\begin{document}
\title{Experimental observation of the longitudinal plasma excitation in intrinsic Josephson junctions
}
\author{A. Irie$^{1}$, Yu.M.Shukrinov$^{2}$, and G. Oya$^{1}$}

\address{$^1$ Department of Electrical and Electronic Systems Engineering, Utsunomiya University, 7-1-2 Yoto, Utsunomiya 321-8585,
Japan}
\address{$^2$ Bogoliubov Laboratory of Theoretical Physics, Joint Institute for Nuclear Research,
Dubna, Moscow Region, 141980, Russia}

\ead{iriea@cc.utsunomiya-u.ac.jp}

\begin{abstract}

We have investigated the current-voltage characteristics (IVCs) of intrinsic Josephson junctions (IJJs).
Recently, it is predicted that the longitudinal plasma wave can be excited by the parametric resonance in IJJs.
Such an excitation induces a singularity called as breakpoint region around switch back region in the IVC. 
We have succeeded in the observation of the breakpoint region in the IVC of the mesa with 5 IJJs at 4.2 K.
Furthermore, it is found that the temperature dependence of the breakpoint current is in agreement with the theoretical prediction.
This suggests that the wave number of the excited plasma wave varies with temperature.
\end{abstract}

\section{Introduction}
Layered high $T_c$ superconductors (HTSCs) represent a stack of alternating superconducting and insulating layers, which form a system of IJJs.\cite{kleiner,oya} 
Coupling between layers leads to the multiple branch structure with hysteresis in the IVC of such stack.
In the system, hysteretic behavior is different from the case of single Josephson junction.
For example, the value of return current depends on from which branch the system has returned to the zero voltage state.
In addition, it has been recently predicted that the system shows a new feature called as breakpoint region (BPR), on the IVC.\cite{prb,prl,sust}
The BPR is characterized by the breakpoint current (BPC) $I_{bp}$ and transition current $I_j$.
The former is a resonance point, at which the longitudinal plasma wave (LPW) is created in the stack with given number and with given distribution of the rotating and oscillating IJJs, while the latter  corresponds to the jumping point from the given branch to the another one (see Fig.~\ref{simulated}(a)).
However, to our knowledge no precise experiment to observe the BPR has been carried out so far.

Furthermore, the problem to determine the value of the coupling parameter in the stack of the IJJs also attracts much interest, but there are still many questions concerning this topic.
The new opportunity to make this problem clear gives the investigation of the breakpoint in the current-voltage characteristics and temperature dependence of the breakpoint current.

In this paper we present the experimental observation of the BPR as resulting from the longitudinal plasma excitation in IJJs.

\section{Theoretical background}


Physical properties of the IJJs in HTSC might be understood in the framework of the capacitively coupled Josephson junction (CCJJ) model.~\cite{koyama96,matsumoto99,physC1}
Inclusion of the diffusion current (DC) leads to the multiple branch structure close to the equidistant one. 
In such CCJJ+DC model, phase dynamics is described by the system of equations~\cite{machida00,physC2}
\begin{eqnarray}
\frac{d^2}{dt^2}\varphi_{l}=(I-\sin \varphi_{l} -\beta\frac{d\varphi_{l}}{dt})+ \alpha (\sin \varphi_{l+1}+
\sin\varphi_{l-1}- 2\sin\varphi_{l}) \nonumber \\+ \alpha
\beta(\frac{d\varphi_{l+1}}{dt}+\frac{d\varphi_{l-1}}{dt}-2\frac{d\varphi_{l}}{dt}) \label{d-phi-dif}
\end{eqnarray}
for the gauge-invariant phase differences $\varphi_l(t)=
\theta_{l+1}(t)-\theta_{l}(t)-\frac{2e}{\hbar}\int^{l+1}_{l}dz A_{z}(z,t)$  between superconducting layers
(S-layers). 
Here, $\theta_{l}$ is the phase of the order parameter in S-layer $l$, $A_z$ is the vector
potential in the barrier, $I$ is the bias current, $\beta$ and $\alpha$ are the dissipation and coupling parameters, respectively. The CCJJ+DC model is different from the CCJJ model by the last term on the right hand side. This coupled Ohmic dissipation term might be derived by the microscopic
theory~\cite{machida00} or phenomenologically by the inclusion of the diffusion current between layers.\cite{physC2}

\begin{figure}[b]
\begin{minipage}{26pc}
\includegraphics[width=20pc]{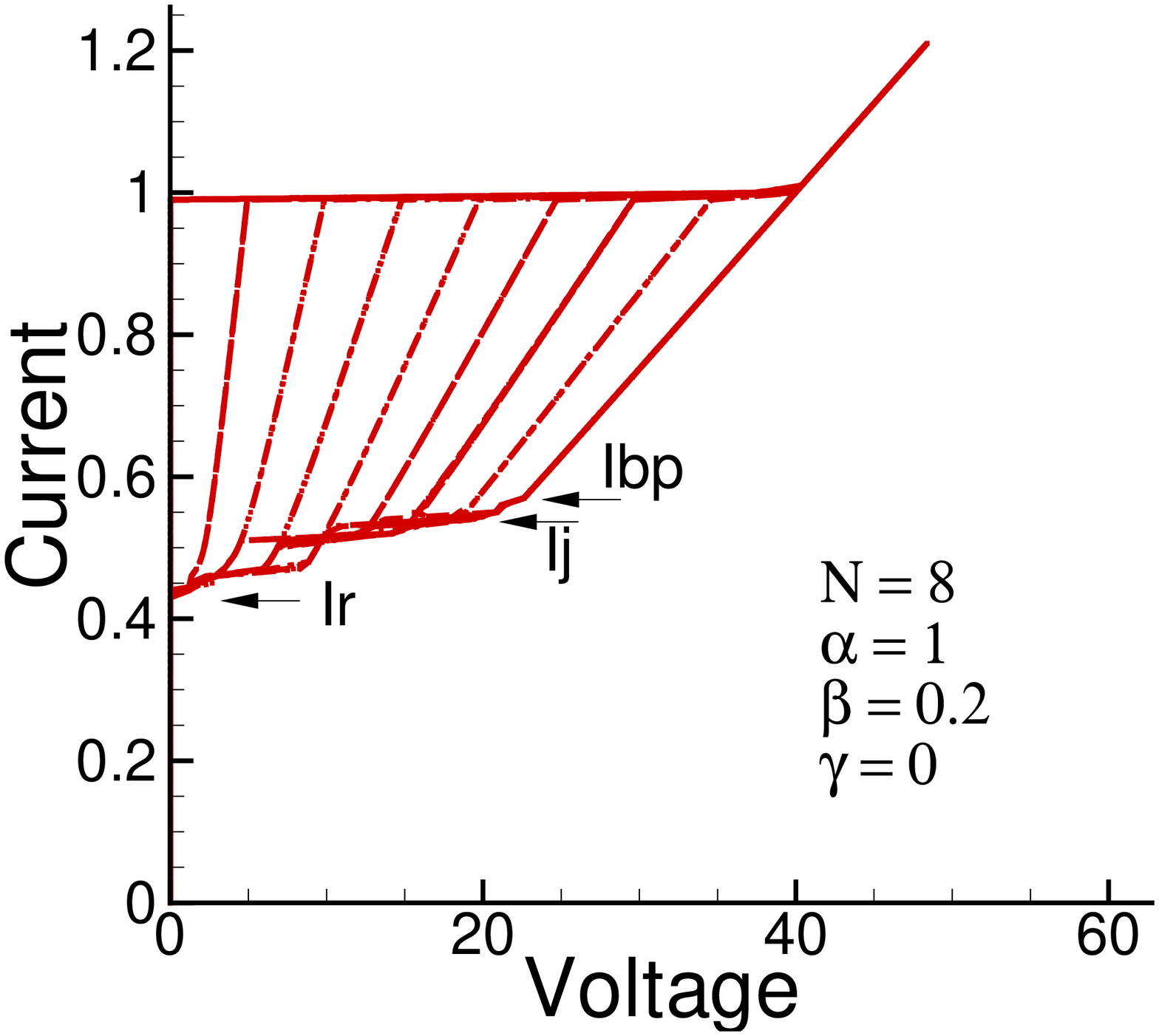}\includegraphics[width=20pc]{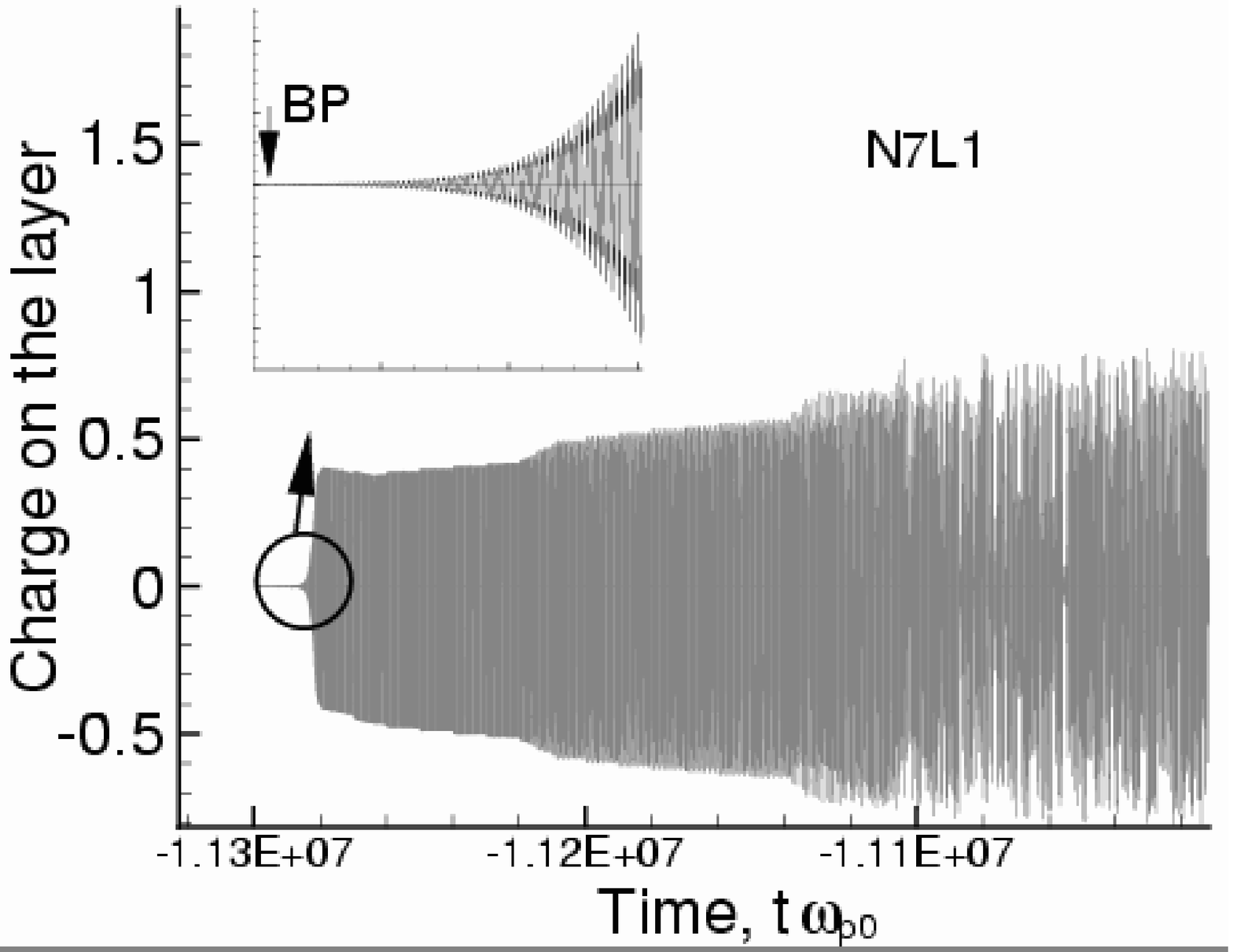}
\end{minipage}
\caption{\label{simulated} (Color online) (a) The simulated IVC of a stack of 8 IJJs for $\alpha = 1$, $\beta =
0.2$ and $\gamma = 0$; (b) Profile of the charge oscillation on the first S-layer in the stack with 7 IJJ. The
inset shows the charge oscillation on the first layer in the beginning of the BPR.}
\end{figure}

Fig.~\ref{simulated}(a) shows the simulation result of the  IVC of a stack of 8 IJJs for 
$\alpha = 1$, $\beta = 0.2$ and  nonperiodic boundary condition $\gamma = 0$, where
$\gamma=s/s_0=s/s_N$ and $s$, $s_0$, $s_N$ are the thicknesses of the middle, first, and last superconducting
layers, respectively.  The IVC demonstrates the equidistant branch structure and a breakpoint on their hysteresis branch.

As was shown in Ref.\cite{koyama96}, the system of equations for the CCJJ model has a solution corresponding to the LPW
propagating along the c axis.
In this case, a frequency of the LPW at $I = 0$ and $\beta = 0$ is $\omega_p(k)=\omega_p
\sqrt{1+2\alpha(1-\cos(k(d+s))}$, where $\omega_p$ is the Josephson plasma frequency, 
$k$ is wave vector of the LPW and $d$ is the thickness of the insulating layer.
The Josephson oscillations, which
frequency  is determined by the voltage value, excite  the LPW by their periodical actions, so that at $\omega_J=2
\omega_p$ the parametric resonance is realized and the LPW with definite wave number is created.~\cite{prb,prl,sust}
Then, the wave number depends on the number of junctions in the stack, boundary conditions, dissipation and coupling parameters.
Using Maxwell equations the charge density $\rho_l$ on the superconducting layer $l$ can be expressed by the voltages $V_{l}$ and $V_{l+1}$ in the neighbor insulating layers.
Solution of the system of dynamical equations for the gauge-invariant phase differences between S-layers gives us the voltages $V_{l}$ in
all junctions in the stack, and it allows us to investigate the time dependence of the charge on each S-layer.

Fig.~\ref{simulated}(b) shows the  profile of the charge oscillation on the first S-layer in the stack with 7 IJJs and the inset shows the charge oscillation on the first layer in the beginning of the BPR.
Here, the time is normalized to $1 / \omega_{p0}$ and the charge is normalized to $ \rho_0 =\varepsilon \varepsilon _0 V_0/r_D^2$, where $r_D$ is Debay screening length.
In this figure, the time is calculated as $t-T_m I_0/\delta I$, where $T_m$ determines time interval for
averaging, $I_0$ is an initial value of the bias current for time dependence recording,  $\delta I$ is a step in
current.
It demonstrates a sharp increase of the oscillation amplitude corresponding
to the creation of the LPW  with a wave vector $k=6\pi/7$.
Therefore, we can say that the observation of the BPR is an indirect evidence of the LPW excitation.

\section{Experimental results}

\subsection{Samples}

\begin{figure}[ht]
\centering\includegraphics[width=110mm]{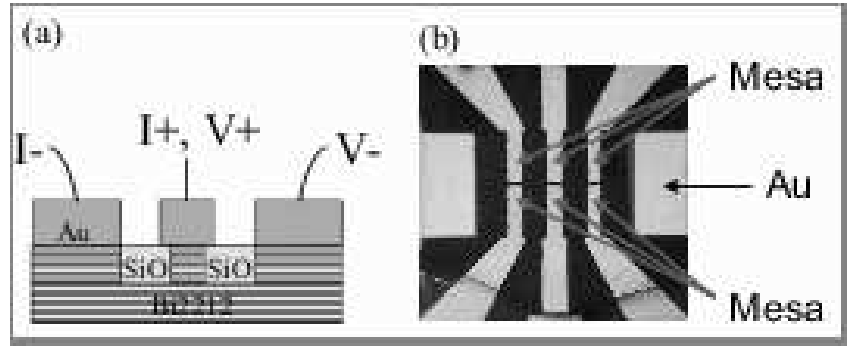}
\caption{\label{sample} (Color online) (a) Schematic view of the fabricated mesa structure and (b) photograph of the sample.}
\end{figure}

Bi$_2$Sr$_2$CaCu$_2$O$_y$ single crystals with critical temperatures $T_c$ of $\sim 85$ K were grown by a
conventional melting method.
The grown single crystals were glued on glass substrates and then cleaved in air to
obtain their fresh surfaces. Subsequently, Au thin films of $20-40$ nm thickness were deposited on the
cleaved surfaces to contact well electrically. Mesas with lateral areas of $9-64~\mu$m$^2$, consisting of  $5 - 20$ intrinsic Josephson junctions, were fabricated by using electron
beam lithography, photolithography and Ar ion milling.
Then the number of IJJs in the mesa was determined by the etching time.
Fig.~\ref{sample} shows schematic view and optical photograph of the fabricated samples.
Their IVC along the $c$-axis direction were measured
using a three terminal method at the temperatures ranging from 4.2 K to $T_c$.

\subsection{Current-voltage characteristics}
\begin{figure}[t]
\centering\includegraphics[width=115mm]{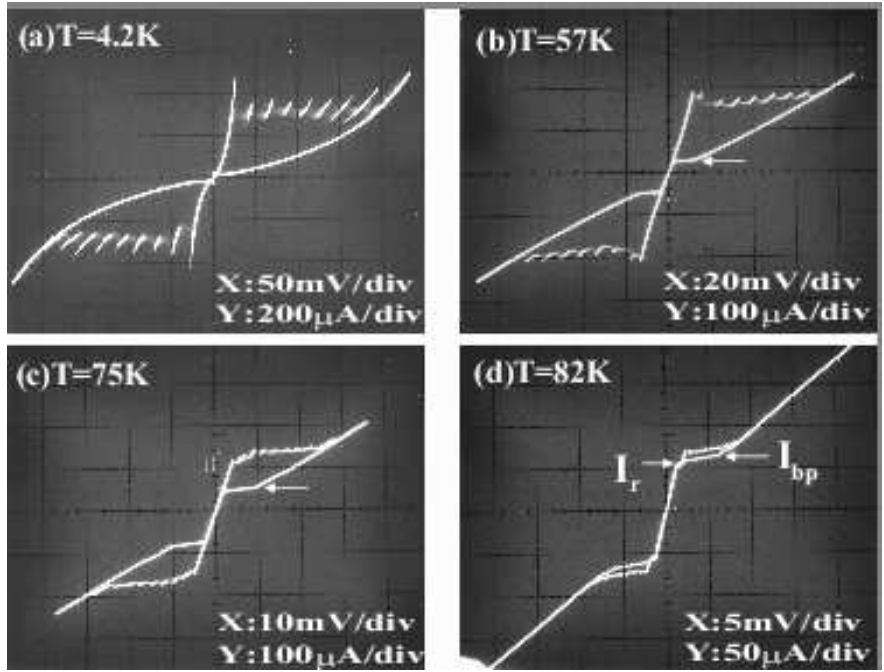}
\caption{\label{iv-T}Typical current- voltage characteristics of a stack of IJJs (samples B) at different temperatures.}
\end{figure}

Fig.~\ref{iv-T} shows the experimental IVCs for the BSCCO stack (sample B) with an area of 24 $\mu$m$^2$ at different temperatures: (a) 4.2 K, (b) 57 K, (c) 75 K and (d) 82 K.
The stack exhibits a large hysteresis at low temperatures and the IVCs have a multiple branch structure.
From the number of resistive branches it is found that this mesa consists of 8 IJJs.
The critical current at 4.2 K is 460 $\mu$A and the corresponding density is 1920 A/cm$^2$.
With the increase in temperature the hysteresis is decreased and disappears at temperatures near $T_c$.
In addition, the increase in temperature leads to the appearance of the breakpoint, which is indicated by arrow on the outermost branch in the IVC.
Similar behavior were also observed for other samples.

\subsection{Temperature dependence of the breakpoint current}
As shown in Ref.~\cite{prl}, the $I_{bp}$ depends on the values of $\alpha$ and $\beta$. Therefore, it would be interesting to see if the predicted features can be observed in the experiment.
According to the resistively and capacitively shunted junction model the hysteresis is related to the dissipation parameter and hence the decrease in the hysteresis as can be seen in Fig.~\ref{iv-T}, corresponds to the increase in $\beta$ under a constant $\alpha$.
Therefore, the predicted features might manifest themselves on the temperature dependence of the breakpoint current.

\begin{figure} \centering
\includegraphics[height=75mm]{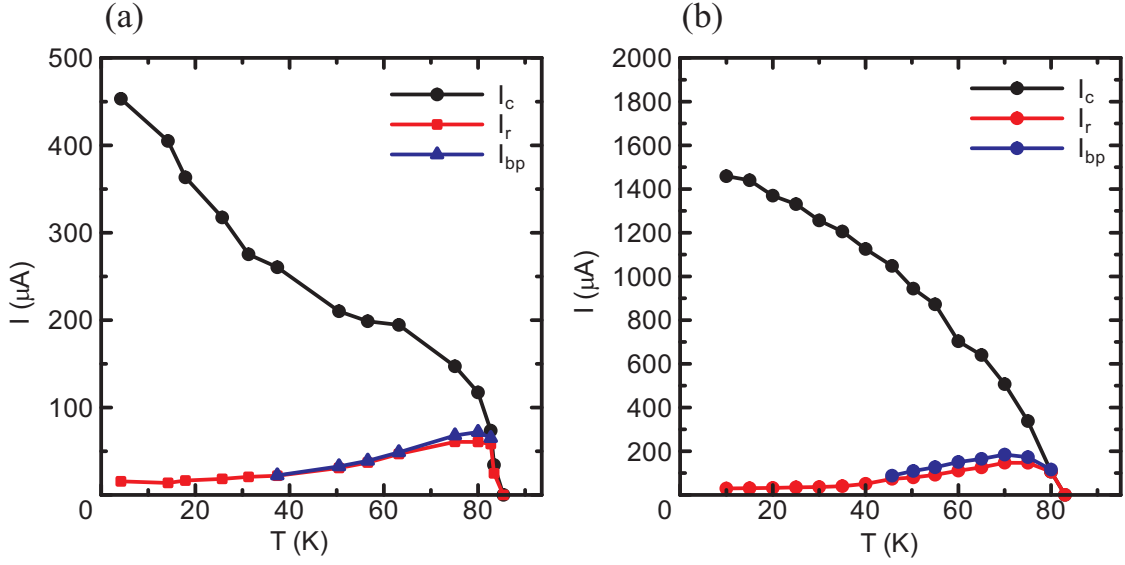}
\caption{\label{bp-T} (Color online) Experimental temperature dependence of $I_c$, $I_r$ and $I_{bp}$ for (a) sample B ($N=8$) and (b) sample No.115 ($N$=30).}
\end{figure}

Fig.~\ref{bp-T}(a) shows the temperature dependence of the critical current $I_c$, the breakpoint current $I_{bp}$ and the return current $I_r$ for sample B.
For this sample the obtained $I_c(T)$ significantly differs from the Ambegaokar-Baratoff dependence, but this may be due to the influence of flux trapping. 
In comparison with the strong $T$ dependence of $I_c$ at low temperature, the $I_r$ hardly depends on $T$ for $T<40$ K, and it gradually increase with $T$ for $T>40$ K.
Such dependence of $I_r$ is typical for IJJ stacks.\cite{oka,kras}
In addition, we note that the $T$ dependence of $I_{bp}$ follows that of $I_r$.
Similar results were also obtained for the samples with larger $I_c$.
An example is shown in Fig.~\ref{bp-T}(b).
It is found that $I_r$ and $I_{bp}$ show essentially the same $T$ dependence as that shown in Fig.~\ref{bp-T}(a).

\begin{figure}[t]
\centering\includegraphics[width=80mm]{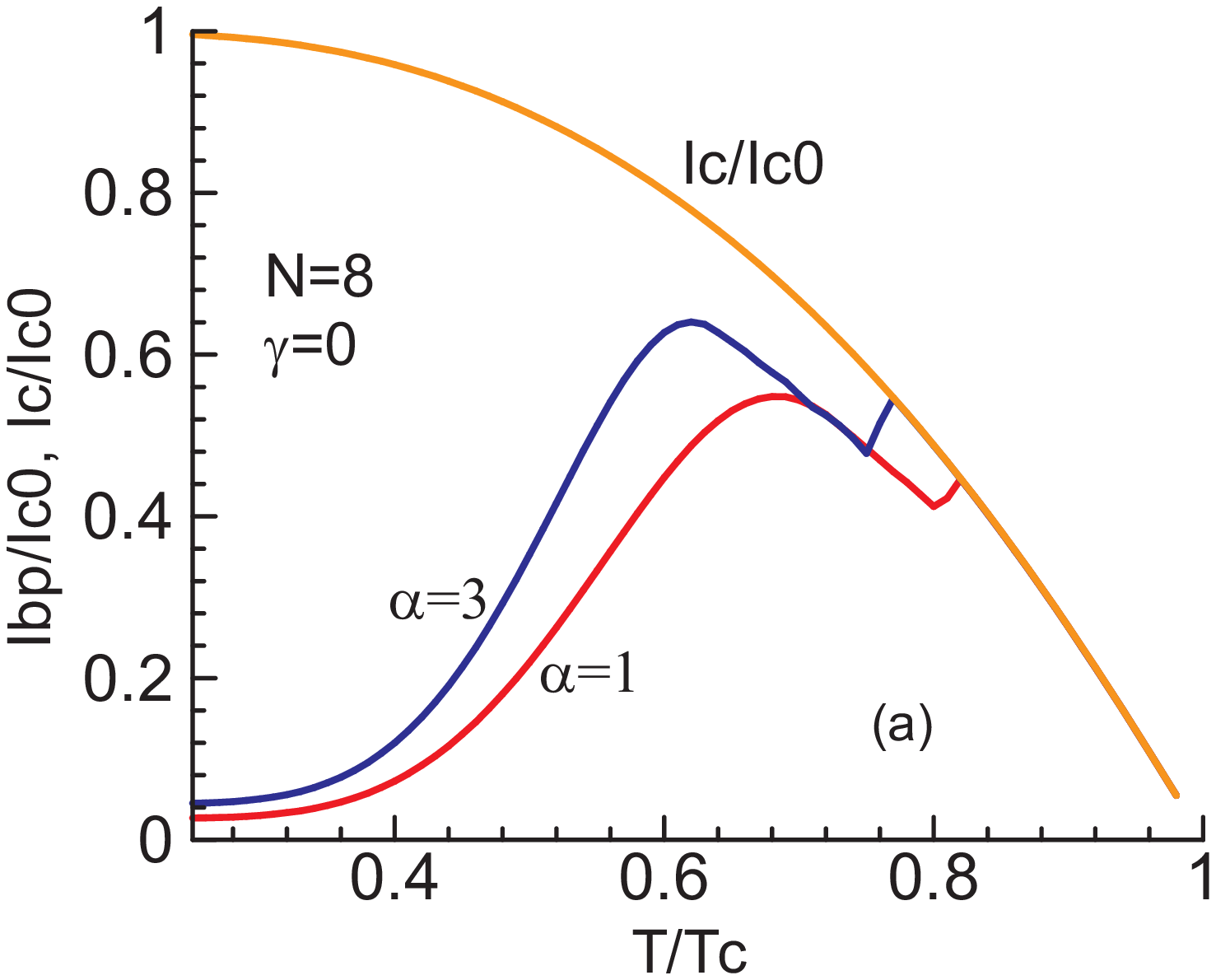}\includegraphics[width=80mm]{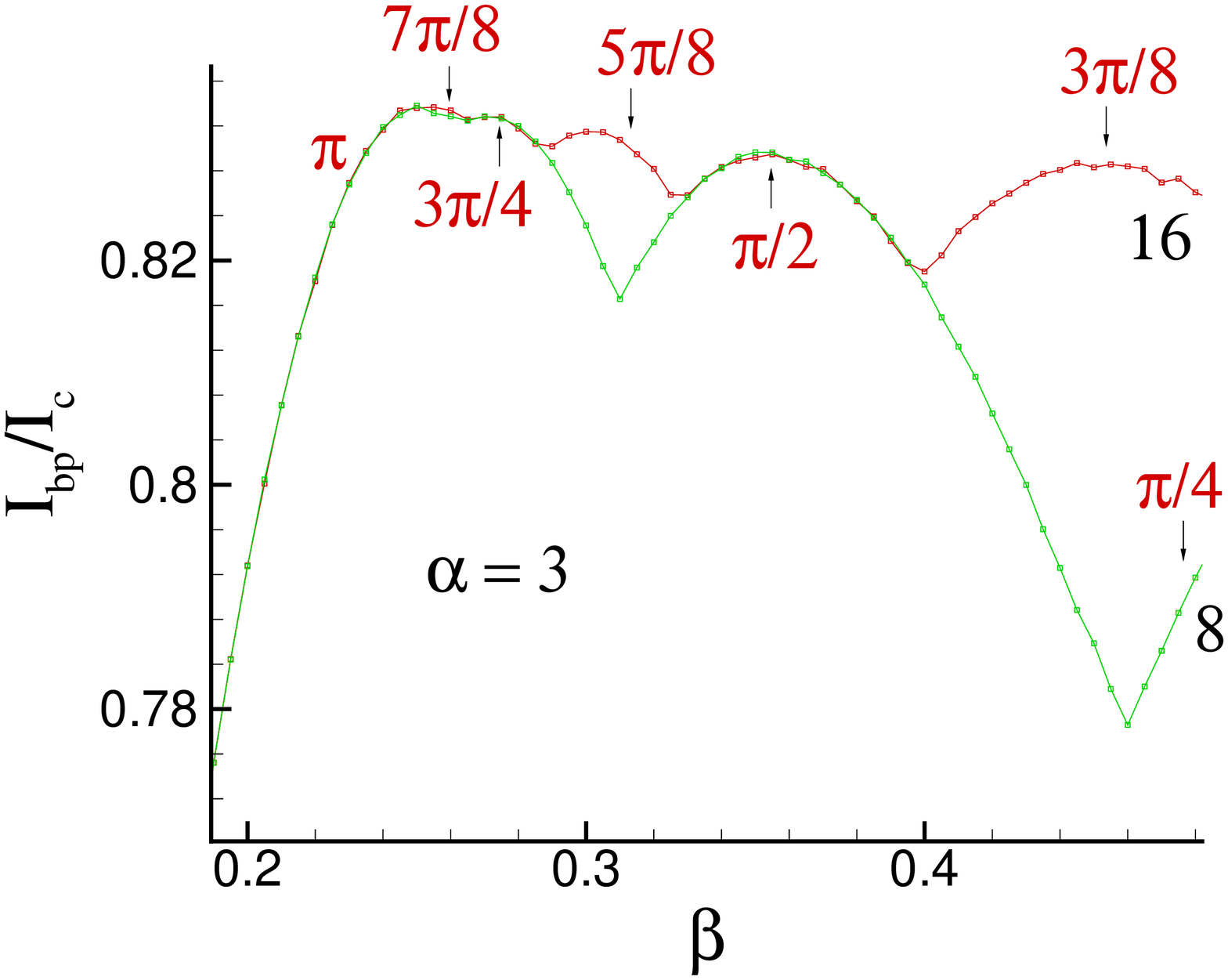}
\caption{\label{T-dep-ibp} (Color online) (a)Temperature dependence of the breakpoint current for the stack with 8 IJJ at different values of coupling parameter ($\alpha=1,~3$). (b) The simulated $\beta$ dependence of $I_{bp}$ for the stacks with 8 and 16 IJJs for $\alpha=3$. In the figure, the wave number of the excited LPW is indicated.}
\end{figure}

In Fig.~\ref{T-dep-ibp}(a), we show the simulated temperature dependence of the breakpoint current for the stack with 8 IJJs at nonperiodic boundary condition $\gamma=0$ for $\alpha=1$ and $3$, which was obtained by using the Ambegaokar-Baratoff dependence of $I_c$.
The details of simulation procedure of the temperature dependence are presented in Ref.\cite{shu-dubna}
It is found that the numerical $I_{bp}(T)$ is qualitatively in agreement with the experimental one.
Furthermore, according to Ref.\cite{prl} the $I_{bp}$ is related to the creation of the LPW with different wave number $k$ at different values of the dissipation parameter $beta$, as shown in Fig.~\ref{T-dep-ibp}(b).
Thus the increase in $I_{bp}$ suggests that the wave number $k$ depends on the temperature.
In order to clear this point, however, a further experiment would be required.

\subsection{Size dependence of the breakpoint current}
\begin{figure}[ht]
\centering\includegraphics[width=150mm]{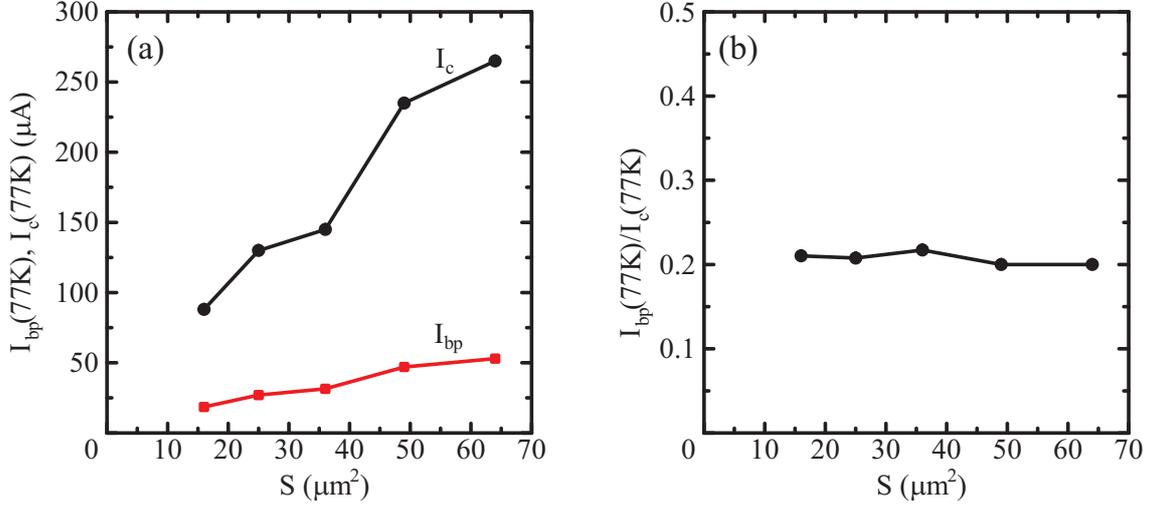}
\caption{\label{S-dep}(a) The $S$ dependence of $I_c$ and $I_{bp}$ of the mesas fabricated on the same crystal, at 77 K. (b) The $I_{bp}/I_c$ ratio as a function of $S$.}
\end{figure}

We have also study the size dependence of the breakpoint current. 
Fig.~\ref{S-dep}(a) shows the $I_c$ and $I_{bp}$ of the mesas, which were fabricated on the same crystal, at 77 K as a function of the mesa area $S$. 
One can see that the $I_c$ and $I_{bp}$ increase lineally with $S$.
The linear increase in $I_c$ indicates that these mesas have almost same critical current density at 77 K : they have a uniform $\varphi$ without the in-plane distribution.
Thus, we attribute the increase in $I_{bp}$ with $S$ to the increase in $I_c$ and hence can say that the breakpoint current is an intrinsic property of IJJs because the value of $I_{bp}/I_c$ is independent of $S$ as shown in Fig.~\ref{S-dep}(b) but depends on $T$.

\subsection{Manifestation of the breakpoint region}
\begin{figure}[ht]
\centering\includegraphics[width=120mm]{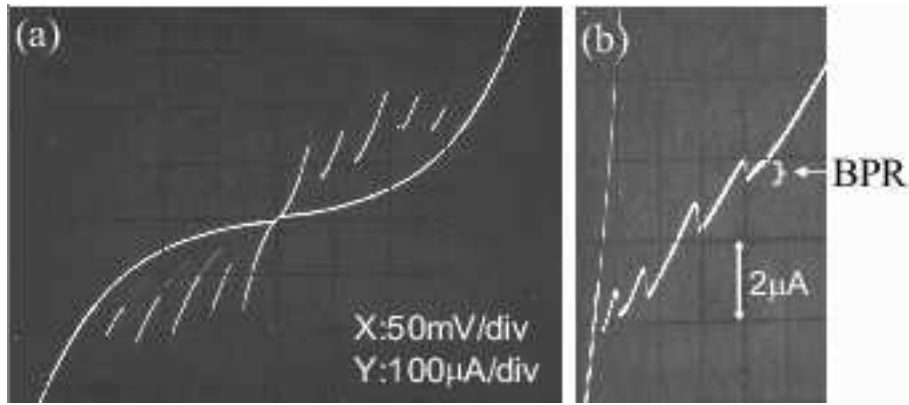}
\caption{\label{bpr} The IVC of the mesa with 5 IJJs at 4.2 K. (a) Overall IVC. (b) IVC in the low bias region.}
\end{figure}

It is usually difficult to observe the BPR on the IVC because the width of BPR, defined as the difference between $I_{bp}$ and $I_j$ is very small and hence in the most case, the $I_{bp}$ coincides with $I_j$ due to thermal fluctuation.
As mentioned above, however, the observation of the BPR would strongly support the excitation of the LPW in IJJ stack.
We succeeded in the observation of the BPR by carefully measuring the IVC. 
The observed BPR is shown in Fig.~\ref{bpr}.
The temperature is 4.2 K and this mesa has an area of $25~\mu$m$^2$ and 5 IJJs.
In Fig.~\ref{bpr}(b), we can see clear BPR on the outermost branch.
The width of the BPR is about 0.5 $\mu$A, and it is found to be considerably smaller compared with the $I_c$ value of $240~\mu$A.

\section{Conclusion}
The current-voltage characteristics of intrinsic Josephson junctions have been investigated.
Clear breakpoint and breakpoint region have been observed on the current-voltage characteristics.
The temperature dependence of the breakpoint current is qualitatively in agreement with theoretical prediction.
Our results imply that the longitudinal plasma wave can be excited in the intrinsic Josephson junctions.

\end{document}